# Variational study of conduction in doped Mott insulator in terms of kinetic energy

H. Yokoyama[a*], S. Tamura[a], K. Kobayashi[b], M. Ogata[c]

[a] *Department of Physics, Tohoku University, Aoba-ku, Sendai 980-8578, Japan*
[b] *Department of Natural Science, Chiba Institute of Technology, Shibazono, Narashino, 275-0023, Japan*
[c] *Department of Physics, University of Tokyo, Bunkyo-ku, Tokyo, 113-0033, Japan*

**Abstract**

The kinetic features of holes (electrons) in the Hubbard model are studied using a variational Monte Carlo method. In doped Mott insulators ($U > U_{\text{co}}$), holes are classified, in the sense of conduction, into two categories: (i) The holes appearing in doublon-holon pairs are localized, encourage singlet pair formation, and gain the energy to induce a superconducting transition. (ii) The doped holes become carriers, make the system itinerant (superconducting or metallic), but lose the energy in the transition. These two roles are distinguished in correlated superconductivity. This feature is common to antiferromagnetic transitions at large $U/t$.

## 1. Introduction

For the relevance to high-$T_\text{c}$ cuprates, the Hubbard model on the square lattice should be thoroughly studied [1]. It is crucial to clarify how the properties of this model evolve as the correlation strength $U/t$ increases, in ascertaining the applicability and limitation of weak- and strong-correlation theories. To this end, one has to adopt non-perturbative approaches that continuously connect the two limits ($U/t \to 0$ and $\infty$) with reliable accuracy. Among such approaches, the variational Monte Carlo (VMC) method has a merit in sufficient wave-number resolution, i.e., applicable system sizes, on realistic lattice systems. By applying a VMC method to a *d*-wave superconducting (SC) state $\Psi_d$ near half filling, it was found that a sharp crossover in SC properties arises at $U_{\text{co}} \approx W$ ($W$: bandwidth) [2,3], which is smoothly linked to the Mott transition point $U_\text{c}$ at half filling [2,4], as $\delta \to 0$ ($\delta$: doping rate). For $U < U_{\text{co}}$, the strength of superconductivity (SC) is very weak and $\Psi_d$ has features common to the conventional *d*-wave BCS superconductors (SC), whereas, for $U > U_{\text{co}}$, the SC as a doped Mott insulator (DMI) is robust and driven by the reduction in kinetic energy, and has various unconventional features, which are, however, consistent with the behaviour of cuprates. This result casts a serious doubt on the weak correlation theories for applying them to the realistic correlation strength of cuprates. In this work, as an extension of the preceding study [3], we corroborate this crossover by studying the gain in kinetic energy $E_t$, and show that, in the DMI regime ($U > U_{\text{co}}$), there are two distinct processes in the SC mechanism: pair formation and conduction of carriers. The gain in $E_t$ in the SC transition wholly originates in the former process, and the latter always loses $E_t$, which challenges the common understanding of kinetic-energy-driven SC. Most features discussed here for SC are shared with the antiferromagnetic (AF) state ($\Psi_{AF}$) and transition.

## 2. Formulation

We consider the standard Hubbard model on the square lattice with only nearest-neighbour hopping near half filling,

$$H = H_t + H_U = -t \sum_{\langle i,j \rangle \sigma} \left( c_{i\sigma}^+ c_{j\sigma} + \text{H.c.} \right) + U \sum_j d_j . \tag{1}$$

with $d_j = n_{j\uparrow} n_{j\downarrow}$. The addition of hopping terms to further sites makes no qualitative difference in the present subject. To this model, we apply variational many-body wave functions of the form, $\Psi = P\Phi$, according to Ref. [3]. Here, $\Phi$ is a one-body (Hartree-Fock) part composed of Slater determinants ($\Phi_d$: $d$-wave BCS state, $\Phi_{AF}$: AF state, $\Phi_F$: Fermi sea), and $P = P_Q P_G$ is the many-body part composed of the onsite Gutzwiller projector $P_G = \prod_j [1 - (1-g)d_j]$ and the doublon-holon (D-H) binding correlation $P_Q = \prod_j (1 - Q_j)$ [5] with

$$Q_j = \mu \left[ d_j \prod_\tau (1 - h_{j+\tau}) + h_j \prod_\tau (1 - d_{j+\tau}) \right] \text{ for } \delta = 0, \quad Q_j = \mu \left[ d_j \prod_\tau (1 - h_{j+\tau}) \right] \text{ for } \delta > 0. \tag{2}$$

A doublon (holon) denotes a doubly occupied (empty) site, $h_j = (1 - n_{j\uparrow})(1 - n_{j\downarrow})$, and $\tau$ in the products runs over the four nearest neighbour sites of site $j$. A D-H binding factor such as $P_Q$ is indispensable for describing a Mott transition and DMI. In $P$, we have two variational parameters $g$ and $\mu$, which are optimized simultaneously with the others in $\Phi$: $\Delta_d$ (singlet gap) and $\zeta$ (chemical potential) in $\Phi_d$, or $\Delta_{AF}$ (AF gap) in $\Phi_{AF}$. To minimize $\langle \Psi | H | \Psi \rangle / \langle \Psi | \Psi \rangle$ and calculate quantities with the optimized parameters, we follow the same VMC procedure as in Ref. [3]. We use the systems of $N_s (= L \times L)$ sites and $N$ electrons ($\delta = 1 - N/N_s$).

## 3. Results and discussions

Because the properties typical in the DMI regime are common among $\Psi_d (= P\Phi_d)$, $\Psi_{AF} (= P\Phi_{AF})$, and $\Psi_F (= P\Phi_F)$, we will mainly discuss $\Psi_d$ and $\Psi_F$ in the following. Around the crossover point $U_{co}/t$ in $\Psi_d$, various SC properties undergo substantial changes [2,3]. Among those, here, we take notice of the mechanism of DC conduction; we aim to obtain an intuitive picture of conduction within the scope of kinetic energy $E_t = \langle H_t \rangle / N_s$. Quantitative discussions using direct measures such as the Drude and SC weights are left for other publications [6]. For the present purpose, it is useful to analyze the kinetic energy by dividing it into two components ($E_t = E_d + E_h$), according as the hopping process varies ($E_d$) or does not vary ($E_h$) the number of doublons, as shown in Fig. 1(a). $E_d$ is derived from the hopping processes that creates or destroys D-H pairs and its effects are, with that of $\langle H_U \rangle$, included in the $J$ term (and the three-site terms if any) in the $t$-$J$ model. $E_h$ comes from the direct hopping of holons and doublons, which roughly corresponds to the $t$ term in the $t$-$J$ model. $E_d$ and $E_h$ are easily calculated using a VMC method [7]. In Fig. 1(c), we plot $E_d$ and $E_h$ of $\Psi_d$ as a function of $U/t$ for several $\delta$. The behaviours of the two components definitely change at approximately $U_{co}/t$.

To begin with, we discuss features in the strongly correlated regime ($U > U_{co}$). At half filling, $E_h$ substantially vanishes in the Mott insulating regime ($U > U_c$), while $E_d$ remains finite and behaves proportionally to $-4t^2/U$ ($= -J$). Since the state at $\delta = 0$ is insulating, $E_d$ here does not contribute to current, namely, local charge fluctuation of creating and annihilating D-H pairs are only repeated. They correspond to the large-$\omega$ part in the conductivity $\sigma(\omega)$: transitions between the lower and upper

Hubbard bands. When carriers are doped, this behaviour of $E_d$ ($\propto -J$) is basically unchanged, although its magnitude decreases slowly as $\delta$ increases. This $\delta$ dependence of $E_d$ is caused exclusively by the change in the number of doublons ($N_s d$), as shown in Fig. 2. Thus, the nature of local D-H processes at $\delta = 0$ remains intact for $\delta > 0$, meaning that $E_d$ is not involved in conduction or itinerancy. On the other hand, $E_h$ becomes finite for $\delta > 0$, and is still almost constant for $U > U_{co}$ with the magnitude proportional to $\delta$ or a formula given by the renormalized mean field theory, $C(\delta) = 2\delta/(1+\delta)$ [8], as depicted in Fig. 2. This indicates that the independent motion of doped holons is realized for $U > U_{co}$, although their mass is somewhat heavier than that of the free electrons (Fig. 2). Thus, in DMI, there exist two kinds of holons which play entirely different roles in SC; the holons created as D-H pairs play a role of forming local singlet pairs, whereas the holons introduced by doping work as current carriers, as schematically sketched in Fig. 3(b). Thus, the number of doped holes is equal to the number of carriers, which will make a small quasi Fermi surface like the Fermi arc or hole pocket. This feature is nothing but that of the $t$-$J$ model, is consistent with various experiments on cuprates, and supports the very low superfluid densities [9]. The residual (background) spinons (singly occupied sites) remain localized unless the doped holons collide with them; this fact is possibly related to a recent neutron scattering for Bi2212 [10], which argued that the source of magnetic response in doped cuprates is localized spins.

Now, let us consider this aspect in terms of the gain in kinetic energy. Various studies including Refs. [2,3] argued that a SC transition in DMI is driven by the gain in $E_t$, namely, $\Delta E_t = E_t(\Psi_F) - E_t(\Psi_d) > 0$ for $U > U_{co}$, in contrast to the interaction-energy-driven mechanism in conventional BCS SC's with weak electron correlations. Actually, the behaviour of $\Delta E_t$ is shown with stars in Fig. 4(a) for an underdoped $\delta$. In contrast, the interaction energy $E_U = \langle H_U \rangle / N_s$ is lost in the transition in DMI [2,3]. Here, we calculate $E_d$ and $E_h$ also for $\Psi_F$, and show them in Fig. 1(b). Using them, we divide $\Delta E_t$ similarly into two parts ($\Delta E_d$, $\Delta E_h$), and plot the results also in Fig. 4(a). At half filling, because $E_h$ is zero, the gain in kinetic energy is entirely attributed to the gain in $E_d$. For $\delta > 0$, we find in Fig. 4(a) that the D-H processes ($\Delta E_d$) still gains the energy for any value of $U/t$ as in the half-filled case, whereas the direct hopping of holons ($\Delta E_h$) always lose the energy. Namely, the stabilization of the SC state originates in the formation of local singlet pairs through the creation and annihilation of D-H pairs, whereas the mobility of carriers, which contributes to conduction or superfluidity, destabilizes the state. Note that this result sharply contrasts with the ordinary idea for the kinetic-energy driven SC that the pairing in the SC state enhances the mobility of carriers, and thereby the quasiparticle renormalization factor $Z$ (effective mass $m^*$) is increased (decreased) [2,3]. Meanwhile, the present result for $U > U_{co}$ is consistent with that of a recent study [11] based on a $t$-$J$-type model, in which the loss in the hopping energy (the expectation value of the $t$ term, which broadly corresponds to $E_h$) by pairing is clarified using a diagonalization technique in effectively truncated spaces. The results regarding $E_d$ are also consistent, although the correspondence of the magnetic ($J$) term in the $t$-$J$ model to $E_d$ and $E_U$ here is somewhat complicated [12].

We turn to the weakly correlated regime ($U < U_{co}$). Both $E_d$ and $E_h$ behave as an identical form as $E_d \sim c_d U^2/t + E_d(0)$ and $E_h \sim c_h U^2/t + E_h(0)$, respectively, with $c_d$ and $c_h$ being constants, as shown by the guide lines in Fig. 1(c). This common behaviour indicates that every hopping process contributes to $E_t$ in the same way, irrespective of the neighbouring electron configuration, because the Hubbard gap vanishes. Correspondingly, the D-H binding is ineffective in this regime, so that holons cannot be classified into the two kinds, and all electrons can contribute to conduction, as in Fig. 3(a); this leads to a large (ordinary) Fermi surface with a carrier number of $N$. In this ordinary Fermi liquid or BCS SC, the number of doped holes ($N_s\delta$) is not in agreement with the carrier number (rather $\propto 1-\delta$), in contrast to the experimental results on cuprates.

Up to this point, we have concentrated on the *d*-wave state and the SC transition; similar features are also found for the AF state and AF transition in the regime of DMI, as shown in Fig. 1(d) and Fig. 4(b). In this case, the range of crossover is broad around $U/t = 10$, but the behaviours of $E_d$ and $E_h$ has the evident features of DMI for large $U/t$.

In summary, we have discussed some features in conduction of DMI through the analysis of kinetic energy, and distinguished them from those of a BCS SC and a Fermi liquid. In succeeding studies, we would like to address the transport properties of (doped) Mott insulators more quantitatively.

**Acknowledgements**

The authors thank L. Vidmar and J. Bonča for useful correspondence. This work is supported in part by Grants-in-Aid from the Ministry of Education, Culture, Sports, Science, and Technology, Japan.

**References**

[1] P.W. Anderson, Science, **235** (1987) 1196.
[2] H. Yokoyama, Y. Tanaka, M. Ogata, H. Tsuchiura, J. Phys. Soc. Jpn. **73** (2004) 1119.
[3] H. Yokoayama, M. Ogata, Y. Tanaka, K. Kobayashi, H. Tsuchiura, to be published in J. Phys. Soc. Jpn. (arXiv:1208.1102).
[4] H. Yokoyama, M. Ogata, Y. Tanaka, J. Phys. Soc. Jpn. **75** (2006) 114706.
[5] H. Yokoyama, H. Shiba, J. Phys. Soc. Jpn. **59** (1990) 3669.
[6] For instance, S. Tamura and H. Yokoyama, an article in this volume.
[7] L.F. Tocchio, F. Becca, C. Gros, Phys. Rev. B **83** (2011) 195138.
[8] F.C. Zhang, C.Gros, T.M. Rice, H.Shiba, Supercond. Sci. Technol. **1** (1988) 36.
[9] Y.J. Uemura *et al*, Phys. Rev. Lett. **66** (1991) 2665.
[10] G. Xu *et al*., Nat. Phys. **5** (2009) 642.
[11] L. Vidmar, J. Bonča, Phys. Rev. B **82** (2010) 125121.
[12] P. Wrobel, R. Eder, R. Micnas, J. Phys: Condens. Mat. **15** (2003) 2755; P. Wrobel, R. Eder, P. Fulde, J. Phys: Condens. Mat. **15** (2003) 6599.

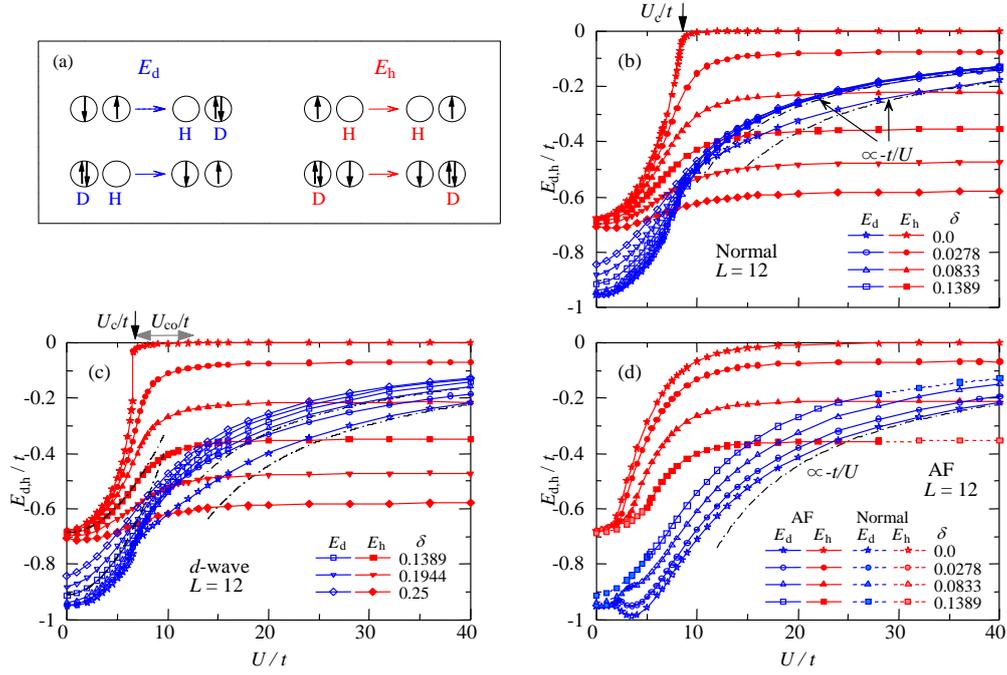

Fig. 1. (a) Hopping processes which contribute to the two kinetic-energy components $E_d$ and $E_h$ are schematically drawn. (b)-(d) $U/t$ dependence of the two kinetic-energy components $E_d$ and $E_h$ of (b) normal, (c) SC [3], and (d) AF states for some $\delta$. The dash-dotted lines for $U > U_{co}$ [$U < U_{co}$ in (c)] are guides of $\sim -t/U$ ($=-J/4t$) [$\sim (U/t)^2 +$ const.]. The arrows on the upper axes in (c) and (d) indicate the Mott transition points ($U_c/t \sim 6.5$-$7$ for $\Psi_d$ and $\sim 8.5$-$9$ for $\Psi_F$ [4]) at half filling and the broad crossover values ($U_{co}/t$) of SC properties for doped cases. In (d), the data of the normal state are plotted in case $\Psi_{AF}$ is not stabilized with respect to $\Psi_F$.

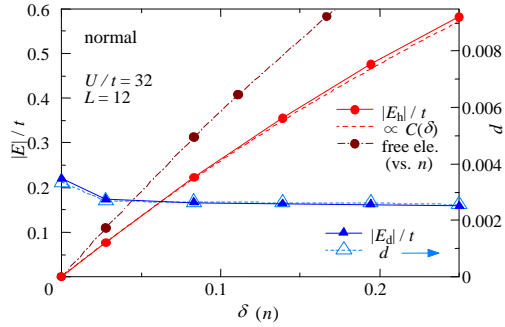

Fig. 2. Absolute values of $E_d$ and $E_h$ for the normal state $\Psi_F$ are plotted as functions of $\delta$ with solid symbols and lines in a strongly correlated case ($U/t = 32$). To compare with $E_d/t$, we add the doublon density $d$ (right axis). To compare with $E_h/t$, we add a guide line $\propto C(\delta) = 2\delta/(1+\delta)$ indicated by the dashed line, and the energy of free electrons by the dash-dotted line as a function of electron density $n$. The features are basically the same as those of the SC state (see Fig. 13(b) in Ref. [3]), as a whole.

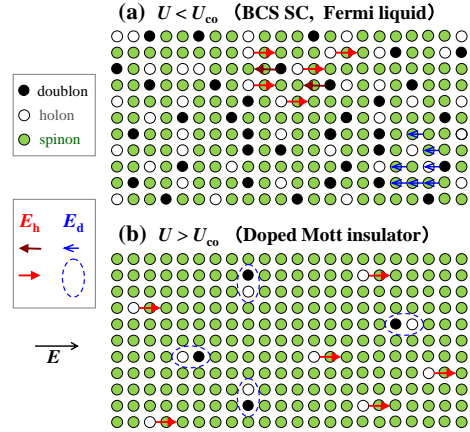

Fig. 3. Microscopic mechanisms in conduction are schematically compared between (a) a BCS-type SC or a Fermi liquid for $U < U_{co}$, and (b) a doped Mott insulator for $U > U_{co}$. Arrows denote the movable directions of particles, if a field $\vec{E}$ is applied. In (a), arrows are omitted except for a part of the sites, for clarity. In (b), the dashed ellipses indicate bound D-H pairs.

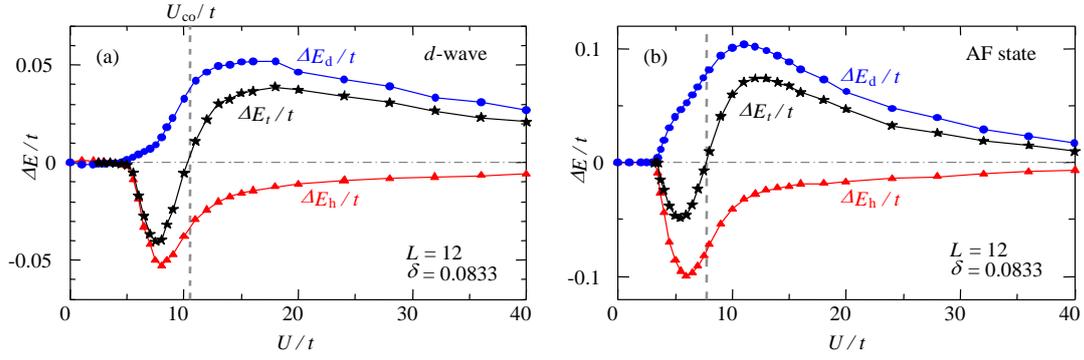

Fig. 4. $U/t$ dependences of kinetic energy gain $\Delta E_t$ (stars) and its components corresponding to $E_d$ (circles) and $E_h$ (triangles), $\Delta E_t = \Delta E_d + \Delta E_h$, for (a) the $d$-wave and (b) AF states for $\delta = 0.0833$. The vertical dashed lines indicate the $U/t$ ($\sim U_{co}/t$) where $\Delta E_t$ changes the sign (loss and gain are reversed). The behaviours (such as $\Delta E_d > 0$ and $\Delta E_h < 0$) are qualitatively the same for other doping rates.